\documentclass{article}
\usepackage{arxiv}
\usepackage{amsthm,amsmath,amssymb}
\usepackage{url}
\usepackage{graphicx}

\title{aPCoA: Covariate Adjusted Principal Coordinates Analysis}
\author{Yushu Shi \thanks{Department of Biostatistics, University of Texas MD Anderson Cancer Center, Houston, TX , 77030, USA.} 
\and Liangliang Zhang \footnotemark[1]
\and Kim-Anh Do \footnotemark[1] \thanks{Equal contribution.} 
\and Christine Peterson \footnotemark[1] \footnotemark[2] \thanks{To whom correspondence should be addressed.}
\and Robert Jenq \thanks{Department of Genomic Medicine, University of Texas MD Anderson Cancer Center, Houston, TX , 77030, USA.} \footnotemark[2] \footnotemark[3]} 

\begin{document}
\maketitle
\begin{abstract}
\textbf{Summary:} In fields such as ecology, microbiology, and genomics, non-Euclidean distances are widely applied to describe pairwise dissimilarity between samples. Given these pairwise distances, principal coordinates analysis (PCoA) is commonly used to construct a visualization of the data. However, confounding covariates can make patterns related to the scientific question of interest difficult to observe. We provide aPCoA as an easy-to-use tool, available as both an R package and a Shiny app, to improve data visualization in this context, enabling enhanced presentation of the effects of interest.\\
\textbf{Availability and implementation:} The R package ``aPCoA" and Shiny app can be accessed at\\
\url{https://github.com/YushuShi/aPCoA.git} and\\
\url{https://biostatistics.mdanderson.org/shinyapps/aPCoA/}\\
\textbf{Contact:} rrjenq@mdanderson.org; cbpeterson@mdanderson.org
\end{abstract}
\section{Introduction}
Non-Euclidean distances, such as Bray-Curtis dissimilarity \cite{BrayCurt57}, unweighted UniFrac distance \cite{LozuKnig05}, and weighted UniFrac distance \cite{LozuHama07} are widely used in fields such as ecology and microbiology  to describe pairwise dissimilarity between samples. In these applications, non-Euclidean distances have critical advantages over  Euclidean distances, such as handling extreme values and incorporating phylogenetic information. Given a non-Euclidean pairwise distance matrix, principal coordinates analysis (PCoA), also known as classic or metric multidimensional scaling (MDS), can allow researchers to visualize variation across samples and potentially identify clusters by projecting the observations into a lower dimension.
  
A long-standing challenge in PCoA visualization is that confounding covariates can mask the effect of the primary covariate. For instance, in a study on the impact of diet on the microbiome, clustering due to site may be more visually prominent than diet if patients are recruited from two different locations. 
Though there have been several methods proposed to adjust for covariates in principal component analysis \cite{ChanDu99, LinYang16},  there are no existing methods to adjust for covariates in PCoA. In this work, we develop a novel visualization approach, aPCoA, which allows adjustment for covariates in creating the PCoA projection, and provide  easy-to-use R tools implementing this method.

\section{Methods}

In this section, we first review the standard steps in creating a PCoA projection from an $N \times N$ distance matrix $\mathbf{D}$ summarizing the pairwise dissimilarity among the $N$ samples in the data set. We then describe how we modify this approach to incorporate covariate adjustment. The standard steps for PCoA are:

\begin{enumerate}
\item Transform $\mathbf{D}$ to a new matrix $\mathbf{A}=[A_{hi}]$, where $a_{hi}=-1/2D_{hi}^2$.
\item Center $\mathbf{A}$ to get Gower's centered matrix $\mathbf{G} = (\mathbf{I}-\frac{\mathbf{11'}}{N})\mathbf{A}(\mathbf{I}-\frac{\mathbf{11'}}{N})$.
\item Calculate the eigendecomposition of $\mathbf{G}$.
\item Project the $N$ samples into 2 dimensions determined by  the two leading eigenvectors.
\end{enumerate}


If the distances are Euclidean embeddable, there exists an $N \times P$ data matrix $\mathbf{Y}=[Y_1,Y_2,\dots,Y_N]'$, such that Gower's centered matrix can be equivalently calculated from $\mathbf{G}=\mathbf{Y}_C\mathbf{Y}_C'$, where $\mathbf{Y_C}$ is $\mathbf{Y}$ centered by the sample mean \cite{Gowe66,LegeLege12}.
 To construct the aPCoA projection, we adjust for the effect of covariates on Gower's matrix, in a manner similar to MANOVA. The $S$ covariates we want to adjust for can be represented in a $N\times S$ matrix, $\mathbf{X}=[X_1,X_2,\dots,X_N]'$, where each $X_k'$ is an $1\times S$ vector. We use a matrix $\mathbf{E}$ to denote the error term which can not be explained after doing a linear regression on $\mathbf{X}$:

\begin{equation}\mathbf{E}=\mathbf{(I-H)}\mathbf{Y}_C,  \end{equation}
where $\mathbf{H}=\mathbf{X}(\mathbf{X}'\mathbf{X})^{-1}\mathbf{X}'$ is the hat matrix used in linear regression. The error covariance matrix, which is also used in pseudo F statistics calculation \cite{Pan11,ZapaScho06} can be calculated by:

\begin{equation}
\mathbf{\Delta}=\mathbf{EE'}=\mathbf{(I-H)}\mathbf{Y}_C\mathbf{Y}_C'\mathbf{(I-H)}. \label{adjusted}
\end{equation}
For any non-Euclidean distance, if we substitute $\mathbf{Y}_C'\mathbf{Y}_C$ in (\ref{adjusted}) with the corresponding Gower's centered matrix $\mathbf{G}$, we can get the generalized error matrix, which is also the covariate adjusted Gower's centered matrix.

\begin{equation}\mathbf{\Delta}^*=\mathbf{(I-H)G(I-H)}. \end{equation}
 After calculating the eigenvectors and eigenvalues of $\mathbf{\Delta}^*$, we can visualize this covariate adjusted Gower's matrix as in a normal PCoA plot. 
 
 We provide aPCoA as both an R package and Shiny app. The Shiny app allows for the adjustment of one covariate, which can be either continuous or categorical, and provides options for visualization including the plotting of 95\% confidence ellipses and lines linking cluster members to the cluster center. Our R package additionally enables adjustment for multiple covariates.
 
\begin{figure}[!tpb]
	\centerline{\includegraphics[width=0.8\linewidth]{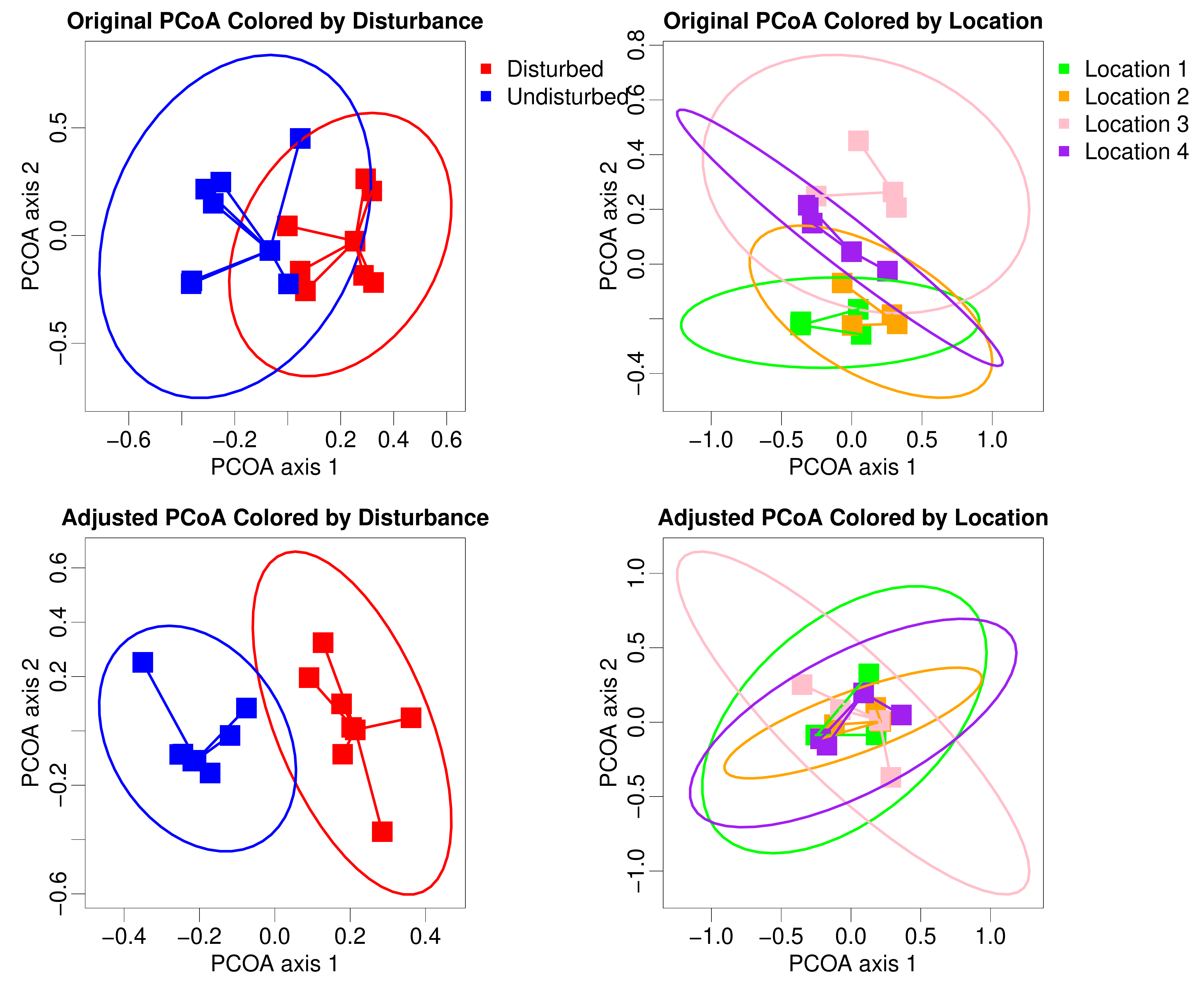}}
	(A) Meiobenthos dataset using the Bray-Curtis dissimilarity\\
	\centerline{\includegraphics[width=0.8\linewidth]{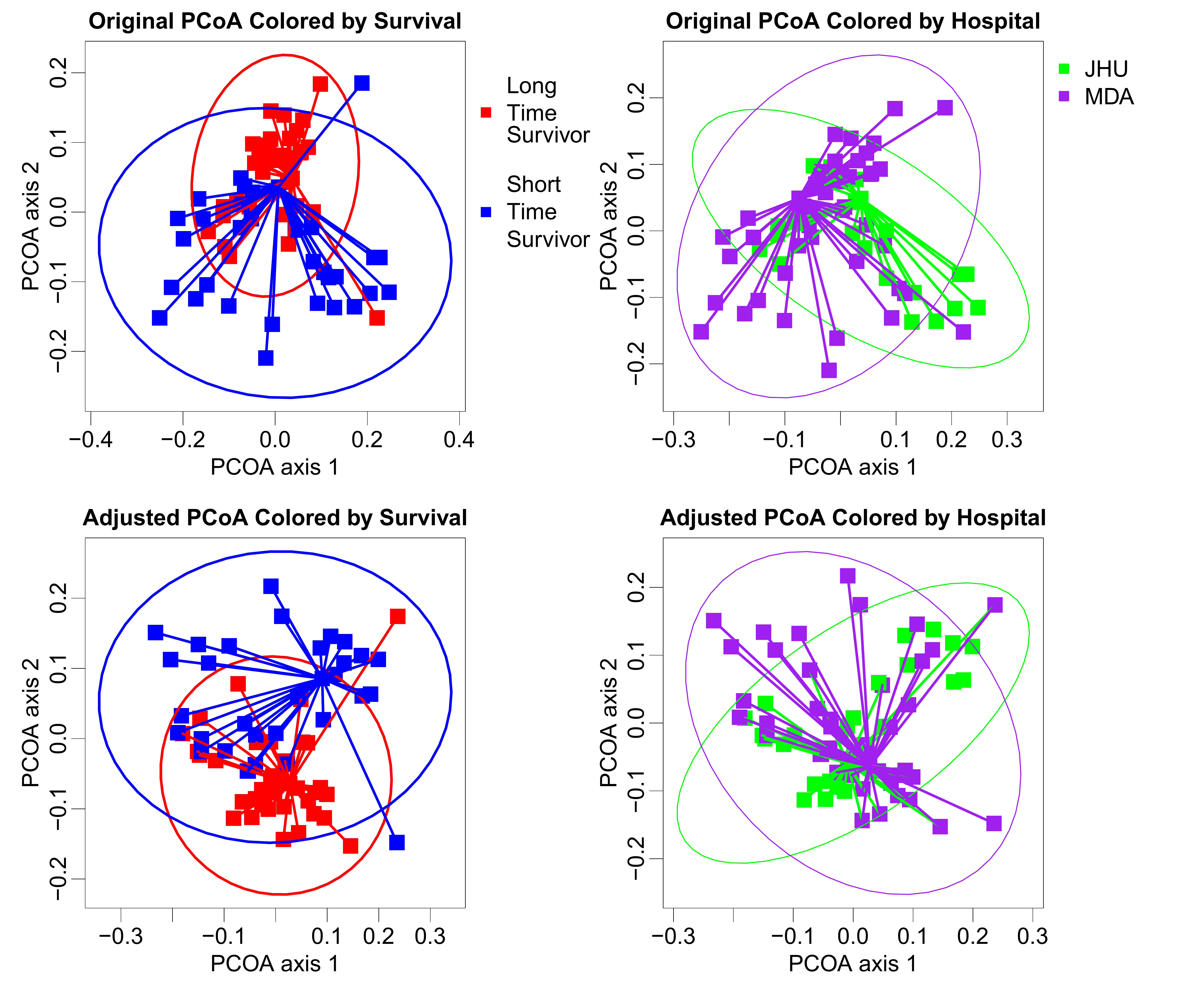}}
	(B) Pancreatic cancer dataset using the weighted UniFrac distance\\
	\caption{\hspace{0pt} Comparison between original PCoA plot and aPCoA plot for two illustrative examples. The position of the points is identical between the left and right columns, with coloring and lines used to illustrate grouping variables.}\label{pic1}
	
\end{figure}
\section{Illustrating Example}
The first illustrating dataset is from a study on the effects of  disturbance from a soldier crab on 56 species of meiobenthos, which are small invertebrates \cite{WangNaum19}.
Eight of the sixteen observations in the data set correspond to crab disturbances. Besides the crab disturbance, there are also four different locations in the study design, where observations from each location are comprised of two disturbed and two undisturbed ones. Here we use the Bray-Curtis dissimilarity, which is commonly used in the ecology field to visualize observations.

The second illustrating dataset is from a two-center pancreatic cancer study \cite{RiquZhan19}, which includes 25 patients from one hospital and 43 patients from another hospital. The investigator compared the tumor microbiota between the 36 long time survivors (LTS) and 32 short time survivors (STS) across study centers. The metric used for visualization is the weighted UniFrac distance, which incorporates both the taxa abundance and phylogenetic relatedness of the bacterial taxa.

As shown in the uppermost panels of Figure \ref{pic1}(A), the original PCoA plot of the meiobenthos dataset is affected by the location, and all  locations are separated from each other. After removing the effect of location using aPCoA, the separation between the disturbed and undisturbed groups becomes more prominent, whereas the separation due to location is less apparent, as shown in the bottom panels of Figure \ref{pic1}(A).

In the pancreatic cancer example, the original PCoA plot with weighted UniFrac distance does not clearly separate the LTS and STS patients due to the confounding effect of hospital site, as shown in the upper part of Figure 1(B). After adjusting for the site effect, the two clusters become more visually separable, as shown in the aPCoA plots provided in the bottom two panels.

\section{Conclusion}
We introduce covariate-adjusted PCoA visualization along with an R implementation, which can help researchers visualize main effects in datasets with strong confounders. We expect our method to be a useful tool for microbiome and ecology research in the future.


\section*{Funding}
KAD is partially supported by MD Anderson Moon Shot Programs, Prostate Cancer SPORE P50CA140388, NIH/NCI CCSG grant P30CA016672, CCTS 5UL1TR000371, and CPRIT RP160693 grants. CBP is partially supported by NIH/NCI CCSG grant P30CA016672 and MD Anderson Moon Shot Programs. RRJ is partially supported by NIH R01 HL124112 and CPRIT RR160089 grants.

\bibliographystyle{unsrt}

\end{document}